\documentclass[5p,times,onecolumn,double-spaced]{elsarticle}

\usepackage[utf8]{inputenc}
\usepackage{graphicx}
\usepackage{subcaption}
\usepackage{siunitx}
\usepackage{multirow}
\usepackage{chemformula}
\usepackage{tabularx}
\usepackage{hyperref}
\usepackage{caption}
\usepackage{booktabs}
\usepackage{tikz}
\usepackage{amsmath}
    \usetikzlibrary{positioning, shapes, snakes}

\address[htu]{College of Physics, Henan Normal University, \textit{Xinxiang 453007}, China}%
\address[imp]{
Institute of Modern Physics, Chinese Academy of Sciences, Lanzhou 730000, China }
\address[HAS]{Institute of Nuclear Science and Technology, Henan Academy of Sciences, \textit{Zhengzhou} 450046, China}
\address[hhu]{College of Energy Engineering, Huanghuai University, \textit{Zhumadian 463000}, China}
\address[CAS]{
 School of Nuclear Science and Technology, University of Chinese Academy of Sciences, Beijing 100049, China}%

\begin{document}

\title{Multiple-models prediction for light neutron-rich isotopes cross section by $Q_g$ systematics in $^{40}$Ar projectile fragmentation reactions}

\author[htu,imp]{Xiao-Bao Wei}
\author[htu]{Hui-Ling Wei}
\author[HAS,htu]{Chun-Wang Ma}
\ead{machunwang@126.com}
\author[htu]{Chun-Yuan Qiao}
\author[htu]{Ya-Fei Guo}
\author[htu]{Jie Pu}
\author[htu]{Kai-Xuan Cheng}
\author[htu]{Yu-Ting Wang}
\author[htu,imp]{Zeng-Xiang Wang}
\author[htu]{Tian-Ren Zhuo}
\author[hhu]{Dan Peng}
\author[imp,CAS]{Shi-Tao Wang}
\author[imp,CAS]{Shu-Wen Tang}
\author[imp,CAS]{Yu-Hong Yu}
\author[imp,CAS]{Xue-Heng Zhang}
\author[imp]{Ya-Zhou Sun}
\author[imp]{Shu-Ya Jin}
\author[imp,CAS]{Guo-Li Zhang}
\author[imp,CAS]{Xuan Jiang}
\author[imp,CAS]{Zhi-Yao Li}
\author[imp]{Ying-Feng Xu}
\author[imp]{Fen-Hua Lu}
\author[imp]{Tuo-Qi Liu}

\begin{abstract}
Precise predictions for nuclei near drip lines are crucial for experiments in new generation of rare isotope facilities.
A multi-models investigation of the $Q_g$ systematics for fragments production cross sections, with $Q_g$ defined as the difference of mass excess (ME) between the projectile ($Z_{p}, A_{p}$) and the fragment ($Z_{f}, A_{f}$) nuclei $Q_{g}=ME(Z_{p}, A_{p})-ME(Z_{f}, A_{f})$, has been performed to verify the model prediction abilities for light neutron-rich isotopes in measured $^{40}$Ar + $^9$Be projectile fragmentation reactions from 57$A$ MeV to 1$A$ GeV. The models used are the FRACS parametrizations and the newly developed Bayesian neural networks (BNN) model. %method
The results show that FRACS, BNN, and $Q_g$ extrapolations are generally consistent, except for fragments near the nuclear mass of the projectile. Additionally, both measured data and model extrapolations provide evidence for a shell closure at $N=$ 16 in fluorine and neon, as well as the disappearance of the traditional magic number $N=$ 20 in neon, sodium and magnesium.
\end{abstract}

%\keywords{Neutron array, Heavy ion collision, Time-of-flight, Symmetry energy, Short-range-correlation}

\maketitle

\section{introduction}
\begin{figure}[htbp]
\includegraphics[width=8.5cm]{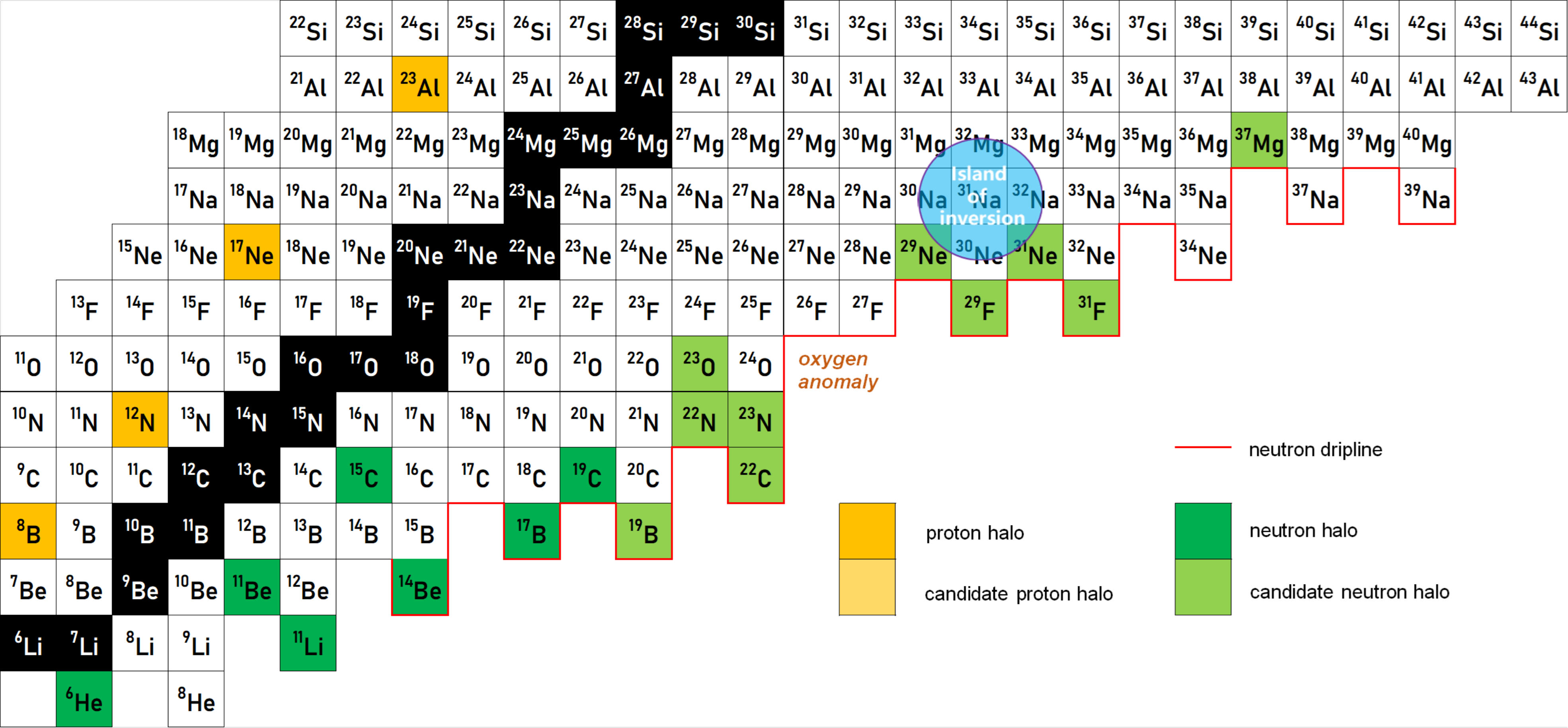}% 
\caption{\label{fig:NC} A section of the nuclear chart showing the exotic structure and unique features exhibit in the light nuclei with extreme neutron-to-protons (n/p) ratios.} % 
\end{figure}

Exploration of new isotopes in the vicinity of the drip line is an important object in nuclear physics. Starting with pioneering experiments at the Lawrence Berkeley Laboratory \cite{40Ar213prl79,48Ca212prl79,Webber90prc}, projectile fragmentation (PF), characterized by its short extraction times and zero-degree emission, has been extensively used to produce and investigate radioactive beams of exotic nuclei \cite{Ma18ppnp,Ma21ppnp}. As indicated in Fig. \ref{fig:NC}, nuclei with extreme neutron-to-proton (n/p) ratios exhibit unique features and behaviors compared to those near the $\beta-$stability line. The existence of exotic structures such as halos and skins has been confirmed in weakly bound nuclear systems \cite{Cai02prc,Tanihata13ppnp,Ma24nst}. In addition, the advance nuclear mass model did not predict an abrupt change in stability from oxygen isotopes to fluorine isotopes, known as the ``oxygen animaly'',  was not predicted by the advance nuclear mass model. This indicates the disappearance of the traditional magic number $N =$ 20 and the emergence of the new shell gap at $N =$ 16 \cite{Ozawa00prl,Stanoiu04prc}. One of the first evidence of significant structure changes was the discovery of enhanced nuclear binding and anomalies in spins, magnetic moments, and the charge radii around heavy sodium $^{31}$Na \cite{Thibault75prc,Huber78prc}. This phenomenon is now understood as a result of the reduced gap between the $sd$ and $pf$ shells, leading to low-lying intruder orbitals to the ground-state configurations of these isotopes, which causes the change of magic numbers and gives rise to large deformation. This region of nuclei ($Z= 10-12$, $N= 20-22$) is called as ``island of inversion'' \cite{Campi75npa,Warburton90prc,Notani02plb,Ahn22prl}. 

Accurate predictions of fragmentation cross sections are crucial in light exotic nuclei experiments, especially in projectile-fragment separators such as LISE++ \cite{LISE04} or MOCADI \cite{MOCADI97}. Many reaction models have been developed to describe the fragment production cross section in PF reactions. Physical models include the statistical abrasion-ablation (SAA) \cite{Brohm94,Gaimard91,Fang00prc} and statistical multifragmentation (SMM) model \cite{Bondorf95pr}, the hybrid macroscopic and microscopic heavy ion phase space exploration model (HIPSE) \cite{Lacroix04prc,Mocko06prc,Mocko08prc}, as well as various versions of microscopic transport models based on the Boltzmann-Uehling-Uhlenbech (BUU) model and the quantum molecular dynamics (QMD) model \cite{Ono04ppnp,Qiao15prc,Ma16nst}. These codes provide valuable insights into the production mechanism and characteristics of near-stable nuclei, but their predictive power diminishes when applied to nuclei far from stability. This limitation arises from the fact that these codes generally rely on the accuracy of the nucleon-nucleon interaction potential and cross sections as well as deexcitation models, which may not be sufficiently precise for light exotic nuclei. 

In addition to the model list above, empirical parameterizations also help predict the fragment cross section in the projectile fragmentation reaction. For example, the EPAX3 formulas \cite{epax3} were constructed based on a large set of reliable experimental data from stable medium and heavy mass projectiles at energies greater than approximately 100$A$ MeV. The accuracy, simplicity  and speed of the analytical formula in calculating the final fragment cross section make it an ideal choice for separator simulation programs \cite{epax2}. However, the assumption that the cross sections are energy and target independent does not align with recent experimental results \cite{Notani07,Kwan12,Zhang12prc}. To address the aforementioned issues, a new set of empirical parametrizations FRACS has been developed \cite{fracs}. Additionally, based on empirical phenomena of fragment production in projectile fragmentation reactions, methods like the scaling formula of fragment cross-sections \cite{scaling19}, the relation of odd-even staggering laws \cite{OES13prc,Mei21cpc}, and the systematical correlation between the fragment cross-sections and the $Q_{g}$ (defined as the difference of mass excesses between the projectile and the fragment $Q_{g}=ME(Z_{p}, A_{p})-ME(Z_{f}, A_{f})$) \cite{Qg07prc,Tarasov09prc,Tarasov13prc}, also shed light on the high-quality predictions of rare fragments with large n/p asymmetry. In recent years, machine learning (ML) has achieved significant success in nuclear physics \cite{Boe.22RevModPhys,He23SCP,He23nst,Pavone23PPCF}, as evidenced by its applications in diverse areas such as fundamental nuclear properties \cite{Niu19prc,Mumpower22prc,Liu24prc}, heavy-ion collisions \cite{Ma20cpc,Qiao21prc,Gao24prc}, effective field theory \cite{Melendez17prc}, shell-model calculations \cite{Yoshida22}, neutron star properties and the nuclear matter equation of state \cite{Drischler20prl}. A recent study proposed a massive learning model based on Bayesian neural networks (BNNs) \cite{Ma22cpc}, demonstrating high precision in reproducing fragment production \cite{Wei22nst,Ma23prc}. Further systematic validation is required to confirm its broader applicability.

This article focuses on the nuclei of the light-neutron-rich nuclear region. Using multiple models for the widely used $^{40}$Ar beam at different bombarding energies, a systematic extrapolation was performed to cross-check the results, providing valuable input for optimizing experimental conditions and guiding the design of experiments in the new generation of rare isotope beam factories. 

\section{Methods}
\label{method}

The main characteristics of the FRACS model are introduced in Sec. \ref{sec.fracs}, following which the $Q_{g}$ systematics is described in Sec. \ref{sec.Qg}, and the BNN model for fragments productions in projectile fragmentation reactions is introduced in Sec. \ref{sec.BNN}

\subsection{FRACS parametrizations}%
\label{sec.fracs}

The FRACS formulas \cite{fracs} are developted from the EPAX2 \cite{epax2} and EPAX3 parameterizations \cite{epax3} incorporating new phenomena in fragments productions, i.e., the incident energy dependence and odd-even staggering (OES) of fragment cross sections. For a specific fragment with mass and charge numbers ($A_f, Z_f$), the production cross section $\sigma(A_f,Z_f)$ in the FRACS formulas is expressed as,
\begin{equation}
\label{sigma}
\sigma(A_f,Z_f)=Y(A_f)Y(Z_{prob}-Z_f)\Delta_{OES}(A_f, Z_f).
\end{equation}
The first term and the second one in the right-hand equation are inherited from EPAX3. $Y(A_f)$ is the isobaric cross section, and it plays crucial roles in determining the fragment production. The incident energy dependence of a fragment is mainly implanted in $Y(A_f)$ of EPAX3, enabling a more accurate representation of how fragment production varies in reactions with different incident energies. The isobaric distribution $Y(Z_{prob}-Z_f)$ is in a Gaussian-like shape about the most probable charge Z$_{prob}$ and the fragment charge $Z_f$ with a given mass $A_f$, except for fragments close to the projectile. Though the first two terms can accurately represent the general smooth trend of fragments cross sections, the OES phenomenon in fragment distributions cannot be described. Mei introduced a new term, $\Delta_{OES}(A_f, Z_f)$, into FRACS. The new FRACS shows a better ability to reproduce the experimental data compared to EPAX3. For a comprehensive description of the FRACS model, readers are referred to a detailed description of FRACS in Ref. \cite{fracs}.

\subsection{$Q_{g}$ systematics}
\label{sec.Qg}
In the two-body transfer reaction, the systematic trends in isotope production cross sections between the ground state $Q-$value ($Q_{gg}$) were well-known \cite{Qgg72PLB,Qgg78PR}. The cross section of fragment can be parameterized in a form of
\begin{eqnarray}\label{eq.Qgg}
\sigma(Z_{f}, A_{f}) = f(Z_{f})\exp[Q_{gg}/T(Z_{f})],
\end{eqnarray}
in which $f(Z_{f})$ is a normalization factor, and $T(Z_{f})$ is an effective temperature adjustable to fit the data. It was used to extrapolate the measured cross sections of fragments and to predict the yields of unobserved ones \cite{Qgg09cpc}. In 2007, Tarasov {\it et al}. firstly apply this relationship to study the projectile fragmentation reaction. As shown in Fig. 4 of Ref. \cite{Qg07prc}, the $Q_{g}$ shows an excellent systematization of the fragment production cross section, with the more neutron-rich side data aligning along an approximately straight line \cite{Qg07prc}, and illustrates that it is more appropriate to describe fragments in projectile fragmentation reactions. The $Q_g$ systematics was also used to extrapolate the extremely low yields of exotic nuclei on the neutron-rich side and to estimate the confidence levels of the unbound isotopes $^{32,33}$F, $^{35,36}$Ne, and $^{38}$Na, helping to confirm the location of the neutron drip line for fluorine and neon in the subsequent discovery of $^{39}$Na \cite{Ahn19prl,Ahn22prl}. As pointed out in Ref. \cite{Tarasov09prc}, the real masses of neutron-rich nuclei are needed in the $Q_g$ method. With predictions of different mass models, the discrepancies in the predicted yields by $Q_{g}$ could potentially help identify missing features in the particular mass model \cite{Tarasov09prc,Tarasov2013}. A sudden change in the slope of the $Q_g$ systematics could indicate alterations of the surface or structure of the nuclear mass \cite{Tarasov09prl,Tarasov13prc}.

\subsection{BNN model}
\label{sec.BNN}
As one type of machine learning, a recent model developed within the framework of BNN technology has shown high precision for rare isotope productions in projectile fragmentation reactions, particular near drip lines \cite{Ma22cpc,Wei22nst,Ma23prc}. The BNN model was constructed based on 6393 measured cross sections for fragments produced in 53 projectile fragmentation reactions. The optimized structure of the BNN model was one input layer and one output layer, and a hidden layer with 46 neural units. Seven parameters were adopted in the input layer, which cover the most important quantities for fragment production [$E, (A_p, Z_p), (A_t, P_t), (A_f, Z_f)$]. The BNN model was announced to be applicable for reactions of incident energy ranging from 40$A$ MeV to 1 GeV/u, projectile nucleus from $^{40}$Ar to $^{208}$Pb \cite{Ma22cpc}, and predictable fragments of $Z>3$ but less than the charge number of the projectile nucleus. The $^{39}$Na produced predicted by BNN in the reaction of $^{48}$Ca + $^{9}$Be at 140$A$ MeV is very similar (0.46 fb) to the experimentally estimated one at 345$A$ MeV at RIKEN (0.5 fb) \cite{Wei22nst}. Omitting the slight incident-energy dependence of fragment production in projectile fragmentation reactions, the BNN prediction has high precision to fragments around the drip lines. The BNN model reported in Ref. \cite{Ma22cpc} was adopted to generate cross sections for fragments in related reactions. 

\begin{figure}[htbp]
\includegraphics{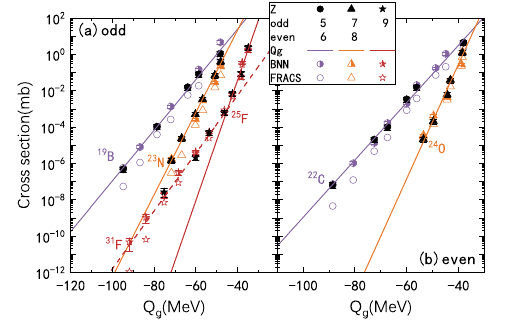}% 
\caption{\label{fig:1000Ar} The cross sections of fragments from boron to fluorine isotopes in the 1$A$ GeV $^{40}$Ar + $^{9}$Be projectile fragmentation reaction measured at GSI (with experimental data taken from \cite{OZAWA00}). The full, open and half-full symbols are for the measured data, the FRACS and BNN predictions, respectively. The results of various isotope chains are depicted using different shapes. The $Q_{g}$ is calculated based on the atomic mass evaluation 2020 (AME2020) \cite{AME2020}. The lines are the fitting results to the measured data.}
\end{figure}

\section{Results and discussion}
With the enhanced neutron richness among natural argon isotopes, $^{40}$Ar is widely utilized as the primary beam for the production of light neutron-rich nuclei at intermediate and relativistic energies because of its ease of handling in an ion source and its abundance. The projectile fragmentation reactions of $^{40}$Ar have been measured with different incident energies in many laboratories around the world, including the GSI in Darmstadt, Germany \cite{OZAWA00}, RIKEN-RIPS, Japan \cite{Notani07}, NSCL, USA \cite{Kwan12}, and HIRFL, China \cite{Zhang12prc}. We will study the $Q_g$ systematics of fragments in the $^{40}$Ar projectile fragmentation reactions based on the above measured data, as well as the FRACS and BNN predictions.

At GSI in Darmstadt, Germany, the $^{40}$Ar beam was accelerated by the SIS synchrotron with energy up to 1$A$ GeV impinged on a Be target \cite{OZAWA00}. The systematic measurements reached the neutron drip line positions of the B, C, N, and O elements, and also reached the drip line position of the F element. The measured data (full symbol) are compared with the predictions of FRACS (open symbols) and BNN (half-full symbols) in Fig. \ref{fig:1000Ar}. The $Q_{g}$ of the fragment is calculated based on the AME2020 atomic mass evaluation version \cite{AME2020}. The systematic variation of all isotopic cross sections as a function of $Q_{g}$ is plotted separately according to their odd or even atomic numbers, respectively. The measured cross section exhibits an excellent linear correlation with $Q_{g}$. The resulting slopes or inverse temperatures parameters obtained by the fitting process are 1/3.12, 1/2.75, 1/1.92 and 1/1.38 MeV for boron, carbon, nitrogen, and fluorine, respectively. In the fluorine isotopic chain, there is an obvious change in slope at $^{25}$ F with $N=$ 16. The four exotic isotopes $^{25-27,29}$F exhibit a reduced slope (shaped by the dashed lines) compared to $^{21, 23-24}$F. The masses of these fluorine isotopes have been measured, and the correlation $\sigma \sim Q_g$ is independent of the models, which provides evidence of $N=$ 16 shell closure in the neutron-rich fluorine isotopes. The BNN model reproduces the measured cross sections for fragments with high precision, resulting in remarkable agreement with the measured $Q_{g}$ systematics. As shown in Fig. \ref{fig:1000Ar}, the FRACS predicts the measured cross sections of oxygen and fluorine well, but the predicted cross sections for lighter nuclei are generally small, which agree with the predictions in Ref. \cite{Wei22nst}. In addition, the predicted results for $^{30,31}$F are inconsistent with BNN and $Q_{g}$. 

\begin{figure*}[htbp]
\includegraphics{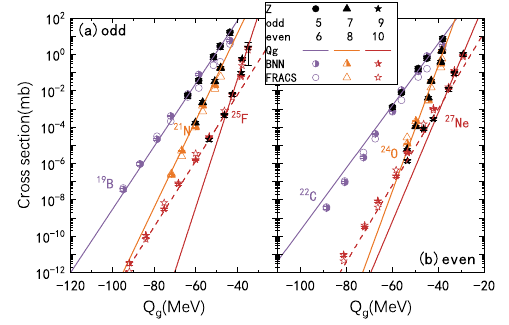}% 
\includegraphics{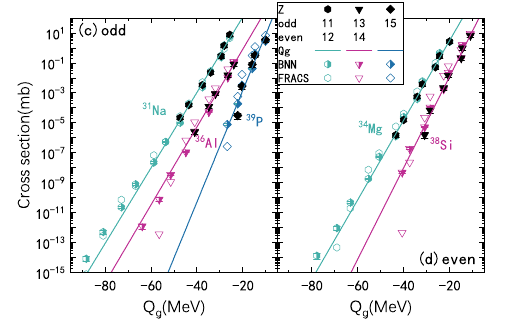}%
\caption{\label{fig:90Ar} Similar to Fig. \ref{fig:1000Ar} but for the 90$A$ MeV $^{40}$Ar + $^{9}$Be measured at RIKEN-RIPS \cite{Notani07}. For odd-isotopes and even isotopes of elements from boron to phosphorus, the $Q_g$ systematics are plotted in (a) and (c), (b) and (d), respectively.}
\end{figure*}

\begin{figure*}[htbp]
\includegraphics{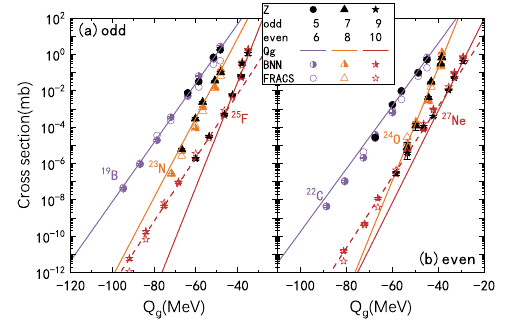}% 
\includegraphics{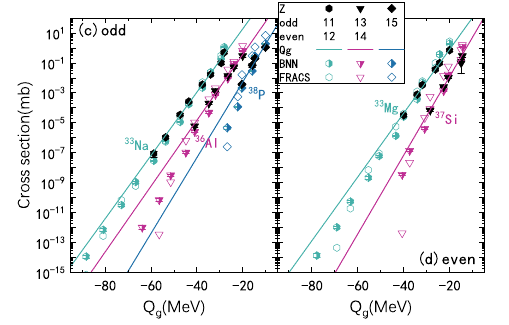}%
\caption{\label{fig:140Ar} Similar to Fig. \ref{fig:1000Ar} but for the 140$A$ MeV $^{40}$Ar + $^{9}$Be reaction measured at NSCL, USA (measured data taken from Ref. \cite{Kwan12}).}
\end{figure*}

At RIKEN-RIPS, the 90$A$ MeV $^{40}$Ar + $^{9}$Be has been measured, and more systematic measurements have been reported for neutron-rich fragments from lithium to phosphorus in 2007 \cite{Notani07}. In Fig. \ref{fig:90Ar}, the $Q_{g}$ systematics for the measured data and the cross sections predicted by FRACS and BNN are plotted. The extrapolation results of the three methods were almost identical, except for those close to the projectile nuclei of $A=$ 40. To be specific, in the fluorine and neon isotopic chains, the dashed lines are fitted by measured data of $^{25,26}$F and $^{27-29}$Ne, respectively, showing evident changes at the slopes of $N=$ 16. Meanwhile, the disappearance of the traditional magic number $N=$ 20 in neon is indicated by the extrapolation of the FRACS and BNN models. The sodium isotopic chain, including $^{25-31}$Na, also contains $N=$ 16, but no significant slope change is observed. Similarly, the traditional magic number $N=$ 20 disappears in the magnesium isotope chain $^{28-34}$Mg. This conclusion is supported by results from the 140$A$ MeV reaction $^{40}$Ar + $^{9}$Be measured by Kwan \textit{et al.} at the National Superconducting Cyclotron Laboratory (NSCL) in 2012 \cite{Kwan12}, as shown in Fig. \ref{fig:140Ar}. Compared to the experiment performed at RIKEN-RIPS, more neutron-rich nuclei of $^{27}$F, $^{30}$Ne, and $^{32,33}$Na were measured. The emergence of the $N=$ 16 shell in fluorine and neon isotopes, and the simultaneous disappearance of the $N=$ 20 shell in neon, sodium and magnesium isotopes are well illustrated by systematics $Q_{g}$. In the extrapolation from $Z=$ 12 to $Z=$ 14, there are large discrepancies between the predictions of BNN and $Q_g$, which exceed an order of magnitude. 

\begin{figure}[htbp]
\includegraphics{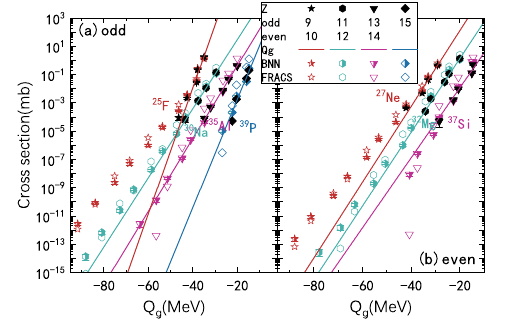}% 
\caption{\label{fig:57Ar} Similar to Fig. \ref{fig:1000Ar} but for fragments in the 57$A$ MeV  $^{40}$Ar + $^{9}$Be reaction measured at HIRFL, China \cite{Zhang12prc}.}
\end{figure}

At the Heavy-Ion Research Facility in Lanzhou (HIRFL), China, 57$A$ MeV $^{40}$Ar + $^{9}$Be was measured to study the mechanisms of projectile fragmentation reactions near the Fermi energy region ($\approx$ 40$A$ MeV) \cite{Zhang12prc}. The production cross sections of fragments from $Z=$ 9 to $Z=$ 20 have been obtained. More complete neutron-rich side data are needed, with results from fluorine to phosphorus shown in Fig. \ref{fig:57Ar}. Since no nucleus with neutron number greater than $N=$ 16 has been detected in fluorine and neon isotope chains, the evidence of the shell $N=$ 16 cannot be seen from the measured fragments. However, the extrapolated results from the theoretical model are in reasonable agreement, showing a significant increase in the cross section of fragments with $N=$ 16.
The FRACS predictions are not applicable to reactions with incident energies below 100$A$ MeV as claimed in Ref. \cite{fracs}, the comparison for the aluminum and phosphorus isotopes in Fig. \ref{fig:57Ar} showing this point.

\begin{figure}[htbp]
\includegraphics[width=8.2cm]{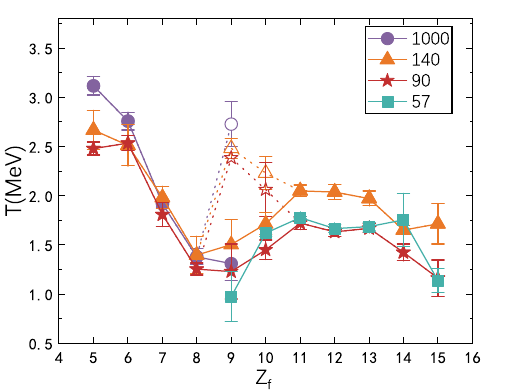}% 
\caption{\label{fig:T} The values of the inverse slope parameter $T$ shown as a function of atomic number ($Z_f$). The different filled (open) symbols connected by solid (dashed) lines represent the inverse slope of solid (dashed) line in four reactions (from Figs. \ref{fig:1000Ar} to \ref{fig:57Ar}).}
\end{figure}

The behavior of the inverse slope parameter (effective temperature $T$), determined by fitting the $Q_g$ systematics in four reactions (from Fig. \ref{fig:1000Ar} to Fig. \ref{fig:57Ar}), is summarized in Fig. \ref{fig:T} as a function of the atomic number ($Z_f$). In general, the extracted $T$ shows a consistent pattern, expect those far from or close to the projectile nucleus's atomic number ($Z$). The values of $T$ in four reactions first gradually decrease, but exhibit a significant increase at the shell location, after which they remain constant for fragments of intermediate charges. Projectile fragmentation is widely accepted as the dominant mechanism for nuclear reactions in the relativistic energy regime. However, as the incident energy decreases, the momentum difference between projectile and target nucleons diminishes. This allows nucleon transfer to become a more significant process. We observe that the temperature of the fragments with charges ranging from 6 to 9 at four energies is very similar, indicating that the formation of these fragments is primarily attributed to the projectile fragmentation mechanism. For heavier fragments at 57 and 90$A$ MeV, the values of $T$ are almost identical except for a deviation at $Z=14$. The trend stabilizes initially and then decreases. This decrease is attributed to the influence of nucleon transfer on fragments approaching the mass of the projectile. While notable errors are observed at $Z=14$ (57$A$ MeV) and $Z=15$ (140$A$ MeV), these deviations do not significantly impact the overall description of the temperature trend observed in the non-relativistic energy regime.

Fragments productions occur not only in projectile fragmentation reactions, but also in fission of superheavy nuclei \cite{fis.95NPA,fis.10PRC}, spallation reactions induced by light particles, including proton \cite{DanP2022JPG}, neutron \cite{Matsumura,Caffee}, $\alpha$ \cite{Vitor,Peter}, and even high-energy $\gamma$ rays \cite{Shibata,Spargin}. It should be very interesting to investigate whether there is $Q_g$ systematics in these reactions, and help the experiments in them for the new generation of rare isotope facilities.

\section{SUMMARY}

In this study, the FRCAS, $Q_g$ and BNN models were adopted to systematically predict neutron-rich nuclei production cross sections in $^{40}$Ar projectile fragmentation reactions at various energies. Such cross-section systematics provide not only a consistency check of the present experiment and theoretical model but also the basis for extrapolation of observed cross sections for future work. Within the $Q_{g}$ framework, both the measured data and the model extrapolations provide evidence for a shell closure at $N=$ 16 in fluorine and neon and the traditional magic number $N=$ 20 disappearance in neon, sodium and magnesium. Analysis of the extraction temperature reveals that lighter fragment formation predominantly arises from fragmentation mechanisms, not only at relativistic energies but also at intermediate energies. 

% \begin{acknowledgments}
% MCW thanks the National Natural Science Foundation of China (NNSFC) under Grant No. 12375123, and the Natural Science Foundation of Henan Province, China under Grant No. 242300421048. SYZ thanks the NNSFC under Grant No. 12305133, and China Postdoctoral Science Foundation under Grant No. 2023M733575.
% \end{acknowledgments}

\end{document}